%% file: main.tex
\documentclass{PoS}
\usepackage[utf8]{inputenc}
\usepackage{float}
\usepackage{subfig}
\usepackage{wrapfig}

\title{Probing the flavor dependence of proton's light-quark sea in the SeaQuest experiment}
\ShortTitle{Probing the flavor dependence of proton's light-quark sea in the SeaQuest experiment}
\author{\speaker{Jason Dove}\thanks{Representing the Fermilab SeaQuest Collaboration.}\\
        Department of Physics, University of Illinois at Urbana-Champaign,\\
        Urbana, Illinois 61801\\
        E-mail: \email{jdove2@illinois.edu}}

\abstract{Surprisingly large flavor asymmetry of the light-quark sea in the proton was reported in deep-inelastic scattering and Drell-Yan experiments.  The Bjorken-$x$ dependence of the $\bar{d}/\bar{u}$ ratio extracted from the Fermilab E866 experiment also revealed an intriguing drop at the highest values of $x$.  The Fermilab E906/SeaQuest experiment was designed to measure $\bar{d}/\bar{u}$ with improved accuracy at high Bjorken-$x$.  By detecting high-mass dimuon events produced in the interaction of 120 GeV proton beam with liquid hydrogen (LH$_2$) and deuterium (LD$_2$) targets, the SeaQuest experiment probes the $\bar{d}/\bar{u}$ ratio up to $x \approx 0.45$.  Data collection has been completed and we report the status of data analysis for SeaQuest.  Two different methods for extracting the LD$_2$/LH$_2$ Drell-Yan cross section ratios for $0.1\leq x\leq 0.45$ are discussed.  These ratios are compared with calculations using various PDFs.}

\FullConference{XXVII International Workshop on Deep-Inelastic Scattering and Related Subjects - DIS2019\\
		8-12 April, 2019\\
		Torino, Italy}

\begin{document}

\maketitle

\input{Sections/FlavorAsymmetry.tex}
\input{Sections/E906.tex}
\input{Sections/Background.tex}
\input{Sections/MassFit.tex}

\bibliographystyle{unsrt}
\bibliography{references}
\end{document}

%% file: Sections/FlavorAsymmetry.tex
\section{Flavor asymmetry of the nucleon sea}
First indications for a flavor asymmetric nucleon sea came from the measurements of the Gottfried sum \cite{Gottfried}

\begin{equation}
    S_G = \int_0^1 (F_2^p(x)-F_2^n(x))\frac{dx}{x} = \frac{1}{3}+\frac{2}{3}\int_0^1[\bar{u}^p(x)-\bar{d}^p(x)]dx.
\end{equation}
For a flavor symmetric sea, $\bar{u}(x) = \bar{d}(x)$, the Gottfried Sum Rule (GSR), $S_G = \frac{1}{3}$ is obtained.  First measurement of $S_G$ at SLAC \cite{Bloom} gave $S_G = 0.28$, suggesting that the GSR is violated.  This early result prompted Field and Feynman \cite{Feynman} to suggest that the Pauli-blocking effect from the $uud$ valence quarks in the proton would inhibit the $\bar{u}u$ sea more than the $\bar{d}d$ sea, leading to a flavor asymmetric $\bar{d} > \bar{u}$ sea.  The most precise measurement of $S_G$ was performed by the NMC Collaboration \cite{Amaudreuz,Arneodo}, giving $S_G = 0.235 \pm 0.026$, establishing the violation of the GSR at a 4$\sigma$ level.

While the violation of the GSR could be explained by a flavor asymmetric sea, it still left open the possibility of pathological behavior of $F_2$ in the unmeasured $x$ region.  A direct and independent experimental tool to probe the flavor asymmetry is to use the proton-induced Drell-Yan process \cite{Ellis}.  By measuring the $\sigma(p+d)/\sigma(p+p)$ Drell-Yan cross section ratio on hydrogen and deuterium targets, the $\bar{d}(x)/\bar{u}(x)$ ratio can be extracted as a function of $x$.  The NA51 Collaboration at CERN \cite{Baldit} found $\bar{u}/\bar{d} = 0.51 \pm 0.04 \pm 0.05$ at $\langle x \rangle = 0.18$, confirming a surprisingly large asymmetry between the $\bar{u}$ and $\bar{d}$ sea in the proton.

The E866/NuSea experiment at Fermilab aimed at a high statistics measurement of 
$\bar{d}(x)/\bar{u}(x)$ covering a broad range in $x$ than the NA51 experiment.  At the leading-order, the Drell-Yan cross section is given as

\begin{equation}
    \frac{d^2\sigma_{_{DY}}}{dx_1dx_2} = \frac{4\pi\alpha^2}{9sx_1x_2}\sum_i e_i^2[q_i(x_1)\bar{q}_i(x_2)+\bar{q}_i(x_1)q_i(x_2)],
    \label{eq:diffCSR}
\end{equation}
where $q_i(x)$ or $\bar{q}_i(x)$ refers to quark or antiquark parton distributions of flavor $i$ evaluated at momentum fraction $x$ of the hadron ($x_1$ and $x_2$ correspond to beam and target hadrons).  At the forward rapidity region ($x_1 > x_2$), the Drell-Yan cross section ratio is approximately given by 

\begin{equation}
    \frac{\sigma_{_{DY}}(p+d)}{2\sigma_{_{DY}}(p+p)} \approx \frac{1}{2}\left(1+\frac{\bar{d}(x_2)}{\bar{u}(x_2)}\right)
\label{eq:CSRApprox}
\end{equation}
Equation \ref{eq:CSRApprox} shows that the $x$-dependence of $\bar{d}/\bar{u}$ can be obtained from a measurement of the $\sigma_{_{DY}}(p+d)/2\sigma_{_{DY}}(p+p)$ ratio as a function of $x_2$.

\begin{wrapfigure}{R}{0.4\textwidth}
\includegraphics[scale=0.7]{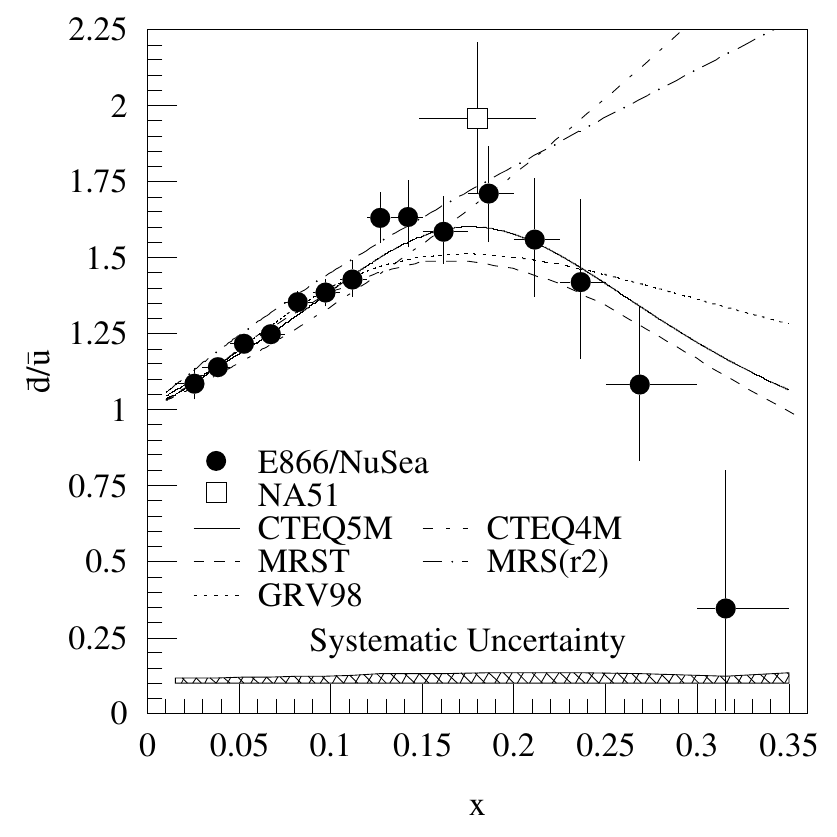}
\caption{NA51 and E866 measurement of $\bar{d}/\bar{u}$ \cite{E866_01}.  Systematic uncertainty is included for the E866 data.  Parametrizations of various PDFs are compared with the data.}
\label{fig:E866Ratio}
\end{wrapfigure}

Figure \ref{fig:E866Ratio} shows the $\bar{d}/\bar{u}$ ratios extracted from the NA51 \cite{Baldit} and E866 \cite{E866_01} experiments over a broad range of $x$.  At the low $x$ region, $\bar{d}/\bar{u}$ approaches 1 as $x\rightarrow 0$, consistent with a symmetric light-quark sea.  As $x$ increases, $\bar{d}/\bar{u}$ rises linearly, reaching a maximum at $x \approx 0.18$, then starting to drop as $x$ further increases.  The striking flavor asymmetry between $\bar{d}$ and $\bar{u}$ has inspired many theoretical models, reviewed in several articles \cite{Speth,Kumano,Garvey,Chang,Geesaman}.  While these models are able to explain the $\bar{d}>\bar{u}$ asymmetry at low $x$, none of the models could reproduce $\bar{d}<\bar{u}$ at the largest $x$.  The E906 experiment was proposed to obtain better precision in the region $x>0.2$.

%% file: Sections/E906.tex
\section{E906/SeaQuest Experiment}
The E906 experiment measures high-mass dimuons produced with a 120 GeV proton beam incident on liquid hydrogen (LH$_2$) and liquid deuterium (LD$_2$) targets.  From Eq. (\ref{eq:diffCSR}), there is an enhancement of the Drell-Yan cross section at lower beam energy compared with E866 due to the $1/s$ factor.  The background due to the open-charm leptonic decay is also expected to be reduced at lower beam energy.  In addition, E906 has a beam intensity of $10^{12}$ protons per second, about 10 times that of E866.

The SeaQuest spectrometer includes two large dipole magnets and four hodoscope/tracking stations.  The first dipole is a solid iron magnet which also serves as a beam dump.  The second magnet has an open aperture allowing the momentum of a charged track to be measured.  Hit information from the hodoscopes was utilized for selecting dimuons originating from the target in an FPGA-based trigger system.  Two 50-cm long LH$_2$ and LD$_2$ targets, together with three solid nuclear targets and an evacuated flask (empty-target), were placed on a movable table so that the targets can be changed between two beam spills.  More details on the E906 spectrometer can be found in \cite{E906_NIM}.

%% file: Sections/Background.tex
\section{E906 Data Analysis}
\begin{wrapfigure}{R}{0.45\textwidth}
\includegraphics[scale = 0.27]{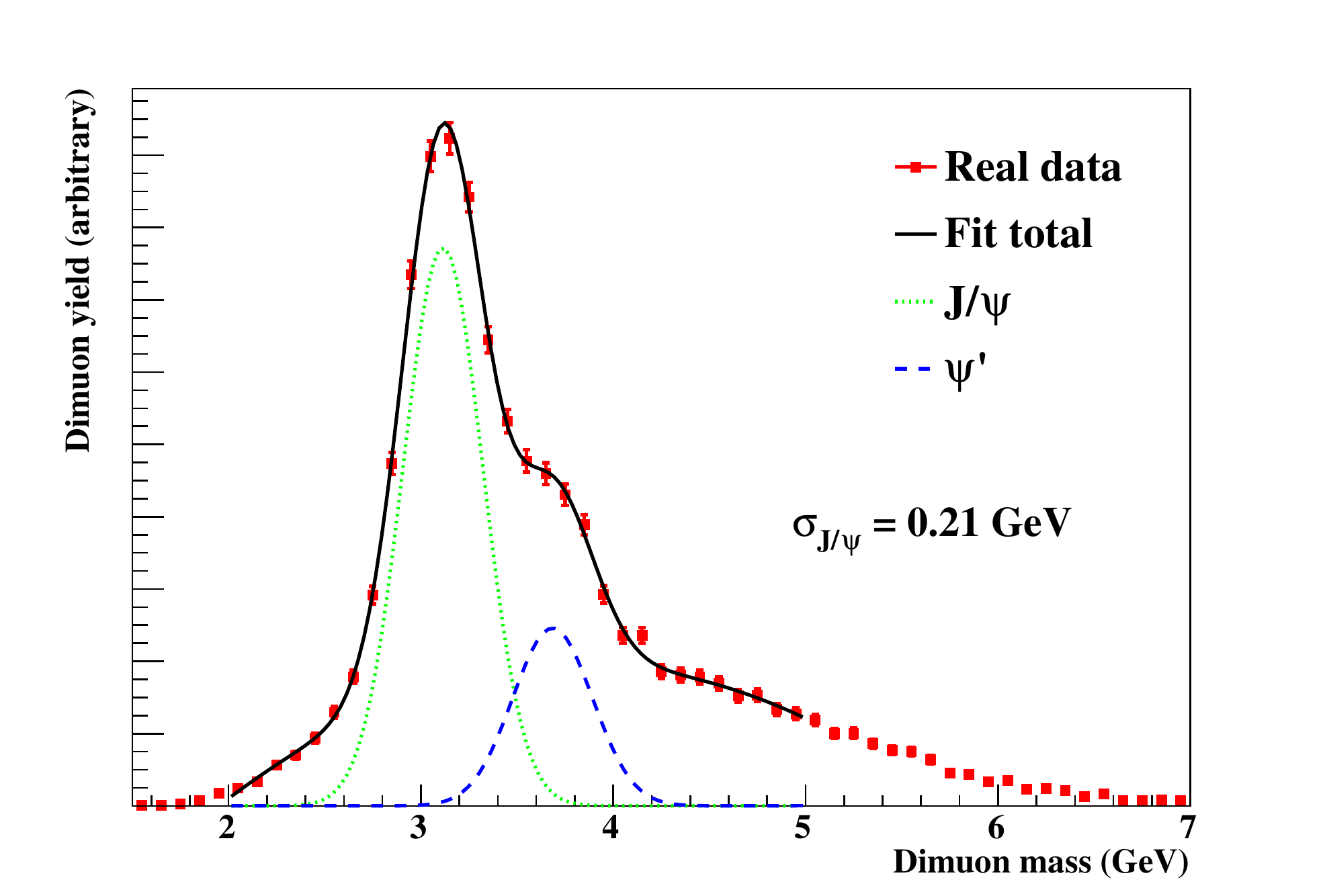}
\caption{Dimuon invariant mass spectrum after applying analysis cuts to select dimuon events from the target \cite{E906_NIM}.  By applying an additional mass cut mass$>4.2$ GeV, the J/$\psi$ and $\psi'$ can be effectively removed.}
\label{fig:MassSpectrum}
\end{wrapfigure}

Data collection for E906 ran from April 2014 to July 2017.  The analysis presented here is performed on the data collected until August 2015, about half of the entire data set.  After applying various analysis cuts to select candidate dimuon events originating from the LD$_2$ target, the dimuon invariant mass spectrum is shown in Fig. \ref{fig:MassSpectrum}.

\noindent The $J/\psi$ peak and the $\psi'$ shoulder are clearly visible.  By selecting events with dimuon mass greater than 4.2 GeV the $J/\psi$ and $\psi'$ events can be effectively removed leaving only Drell-Yan and some background events.

In order to help remove the background from interactions that did not originate from the target, data were also collected on an empty flask.  The empty flask data properly normalized can be used to subtract events which pass the analysis cuts but are in fact originating from either the dump, the flask windows, or the air upstream or downstream of the target.

A major challenge for the E906 experiment was to properly account for the accidental background originating from the coincidence of two independent single-muon tracks.  Due to the high proton beam intensity in E906, the optimal removal of the accidental background is essential.

Two different methods have been used to account for the accidental background in E906.  The first method examines the beam-intensity dependence of the LD$_2$ and LH$_2$ candidate Drell-Yan events.  Because accidental events originate from two independent interactions, the rate for accidental events should have a strong intensity dependence.  By plotting the ratio of deuterium yield over hydrogen yield as a function of incident proton intensity and extrapolating to zero intensity, one can obtain the Drell-Yan cross section ratio without the contamination of accidental events.

\begin{figure}
\centering
\subfloat[]{
\includegraphics[width = 0.45\linewidth]{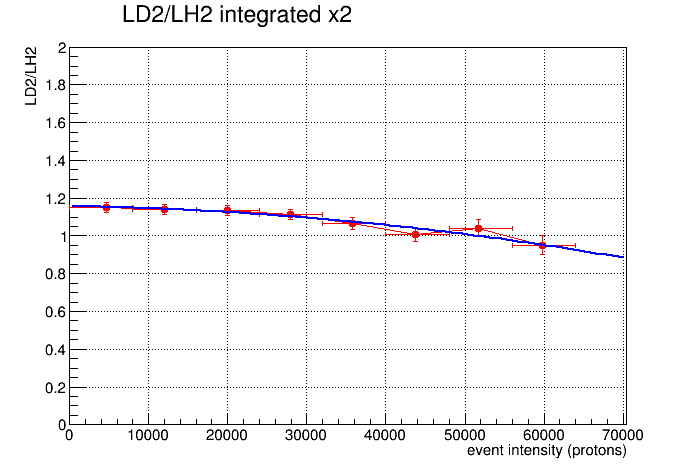}
\label{fig:x2Integrated}
}
\subfloat[]{
\includegraphics[width = 0.45\linewidth]{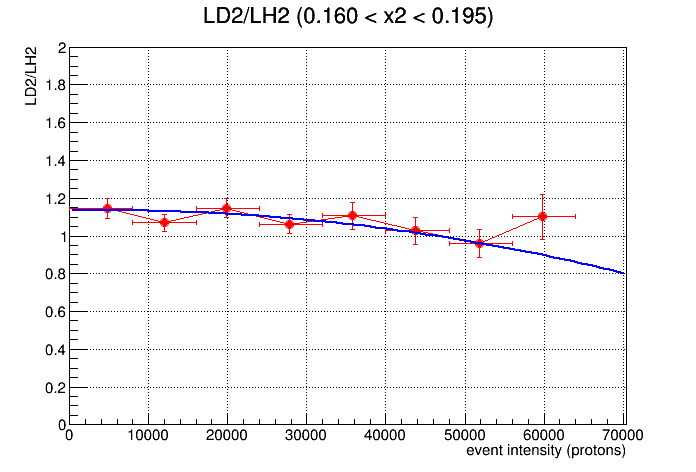}
\label{fig:CSRBin1}
}
\caption{a) Second-order polynomial fit to the beam-intensity dependence of the ratios of deuterium yield to hydrogen yield (LD$_2$/LH$_2$).  The data included all Drell-Yan candidate events.  b) The LD$_2$/LH$_2$ ratios for an $x_2$ bin compared with the curve from a correlated fit to all $x_2$ bins using Eq. (\ref{eq:fitFunction})}

\end{figure}

Fig. \ref{fig:x2Integrated} shows the LD$_2$/LH$_2$ ratios for Drell-Yan candidate events as a function of the proton beam intensity $I$.  From a second-order polynomial fit to the data, shown as the solid curve in Fig. \ref{fig:x2Integrated}, the value of the $\sigma_{pd}/2\sigma_{pp}$ is given as the intercept of the fit.  
In order to extract the $\sigma_{pd}/2\sigma_{pp}$ cross section ratios as a function of $x_2$, the Drell-Yan candidate events are split into seven $x_2$ bins and the intensity dependencies of the LD$_2$/LH$_2$ ratios for all $x_2$ bins are fitted with correlated second-order polynomials as follows:

\begin{equation}
    R_n(I) = P_{0n} + (P_{1}+P_{1}'x)I+(P_{2}+P_{2}'x)I^2
    \label{eq:fitFunction}
\end{equation}
where $n$ ranges from 1 to 7 for the seven $x_2$ bins.  The $P_1'$ and $P_2'$ fitting parameters accommodate the expected smooth variation of the $I$ and $I^2$ dependencies among various $x_2$ bins.  Fig. \ref{fig:CSRBin1} shows the result of this correlated fit to one of the seven $x_2$ bins $(0.16 < x_2 < 0.195)$.

\begin{figure}
\centering
\includegraphics[scale=0.55]{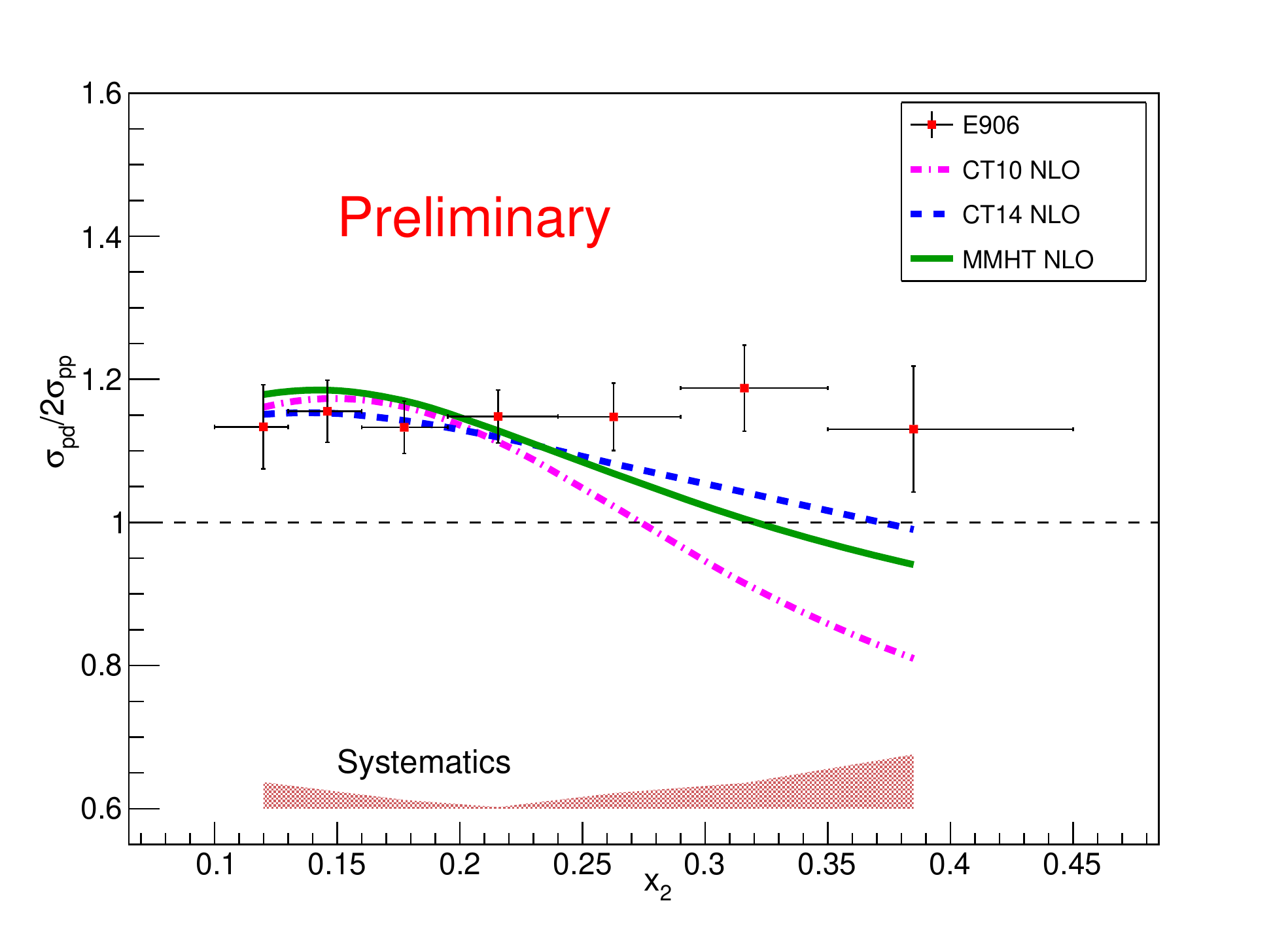}
\caption{Comparison of cross section ratio from E906 data and various PDF fits.  All PDFs and data are in good agreement at low $x$ where the data from previous E866 was well determined.  At higher $x >0.2$ the uncertainty in the E866 data leads to underdetermined $\bar{d}/\bar{u}$ and more variety in the PDFs.  The addition of E906 data at this high $x$ range will provide additional constraints on the PDFs.}
\label{fig:E906_PDFs}
\end{figure}

The $\sigma_{pd}/2\sigma_{pp}$ cross section ratios as a function of $x_2$, obtained from the intercepts $P_{0n}$ in Eq. (\ref{eq:fitFunction}) are shown in Fig. \ref{fig:E906_PDFs}.  Calculations of the Drell-Yan cross section ratios in the Next-to-leading order framework, weighted by the E906 spectrometer acceptance, are shown for three sets of nucleon parton distribution functions (PDFs).  Fig. \ref{fig:E906_PDFs} shows that the E906 result is in good agreement with the calculations for all PDFs at $x<0.24$.  The high statistics of the E866 data at this low $x$ region put stringent constraints on the various PDFs.  At large $x$, the curves begin to disagree with both E906 data and each other.  The large discrepancy between PDFs is indicative of a larger uncertainty in the constraints from E866.  The new data from E906 will clearly provide new constraints for future PDFs at  the large $x$ region

%% file: Sections/MassFit.tex
\begin{figure}[H]
\centering
\includegraphics[width=0.65\textwidth]{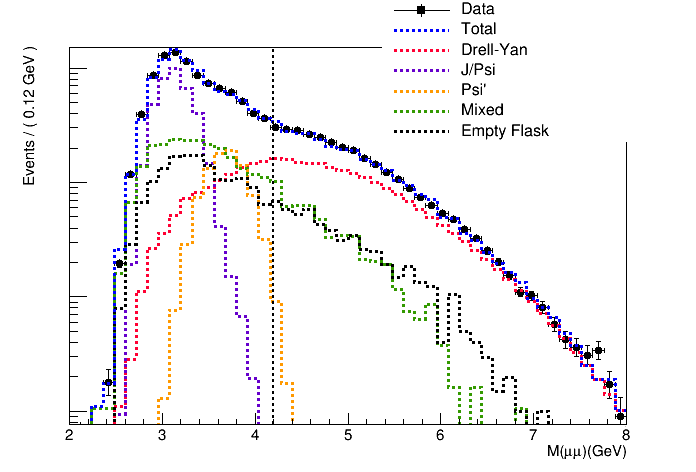}
\caption{Dimuon mass spectrum after analysis cuts designed to suppress dump events and background \cite{E906_NIM}.  Monte Carlos are run to produce expected shapes for $J/\psi$, $\psi'$ and Drell-Yan.  Accidental background is produced by mixing uncorrelated single tracks from data.  Events with the empty flask are normalized by the number of protons incident on targets and are subtracted.  The remaining data is fit to the Monte Carlo and mixed background.  By applying a mass cut $> 4.2$ GeV, the J/$\psi$ and $\psi'$ can be removed.}
\label{fig:MassFit}
\end{figure}

\newline\indent Another method we adopted to account for the accidental background is by fitting the dimuon mass spectrum including various sources of dimouns.  As shown in Fig. \ref{fig:MassFit}, the dimuon mass spectrum is well fitted by a combination of different processes.  The empty flask mass spectrum is taken from data, properly normalized.  The shapes of the $J/\psi$, $\psi'$, and Drell-Yan mass distributions are taken from Monte-Carlo simulations.

The remaining task is to produce an appropriate shape for the accidental events.  Since the accidental events primarily arise from coincidence of single muon tracks, the accidental background is produced by mixing single muon tracks.  Events which contain only a single muon are mixed to ensure the kinematics of the individual tracks is consistent with a single muon, and because the events are separated in time, there is no chance that a true dimuon will be erroneously constructed.  Once the relative normalization between the various components has been fixed, the fit can be projected onto other variables (for example $x_2$) and the backgrounds can be appropriately subtracted and the Drell-Yan yields can be extracted.  Preliminary results from the mass-fit to LD$_2$ and LH$_2$ data show that the LD$_2$/LH$_2$ cross section ratios are consistent with those extracted from the intensity extrapolation method discussed earlier.

